\documentclass[conference]{IEEEtran}

\usepackage{cite, amssymb}
\usepackage{amsmath,amsfonts}
\usepackage{algorithmicx}
\usepackage{graphicx}
\usepackage{textcomp}
\usepackage{tabularx}
\usepackage{array}
\usepackage{xcolor}
\usepackage{hyperref}
\usepackage{float}
\usepackage{placeins}
\usepackage{multirow}
\usepackage{booktabs}
\usepackage{caption}
\usepackage{subcaption}
\usepackage{algorithm}
\usepackage{algpseudocode}
\usepackage{enumitem}

\usepackage{ragged2e}

\begin{document}

\title{Security Assessment and Mitigation Strategies for Large Language Models: A Comprehensive Defensive Framework}

\author{
	\IEEEauthorblockN{Taiwo Onitiju}
	\IEEEauthorblockA{
		School of Computing \\
		University of North Florida \\
		n01578746@unf.edu
	}
\and
	\IEEEauthorblockN{Iman Vakilinia}
	\IEEEauthorblockA{
		School of Computing \\
		University of North Florida \\
		i.vakilinia@unf.edu
	}
}

\maketitle

\begin{abstract}
Large Language Models increasingly power critical infrastructure from healthcare to finance, yet their vulnerability to adversarial manipulation threatens system integrity and user safety. Despite growing deployment, no comprehensive comparative security assessment exists across major LLM architectures, leaving organizations unable to quantify risk or select appropriately secure LLMs for sensitive applications. This research addresses this gap by establishing a standardized vulnerability assessment framework and developing a multi-layered defensive system to protect against identified threats. We systematically evaluate five widely-deployed LLM families GPT-4, GPT-3.5 Turbo, Claude-3 Haiku, LLaMA-2-70B, and Gemini-2.5-pro against 10,000 adversarial prompts spanning six attack categories. Our assessment reveals critical security disparities, with vulnerability rates ranging from 11.9\% to 29.8\%, demonstrating that LLM capability does not correlate with security robustness. To mitigate these risks, we develop a production-ready defensive framework achieving 83\% average detection accuracy with only 5\% false positives. These results demonstrate that systematic security assessment combined with external defensive measures provides a viable path toward safer LLM deployment in production environments.
\end{abstract}


\section{Introduction}
\label{sec:introduction}
The rapid deployment of Large Language Models in production environments has created an urgent need for comprehensive security assessments and robust defensive mechanisms to ensure their safe and reliable operation. As these LLMs process millions of user queries daily in applications ranging from customer service to healthcare decision support, understanding their vulnerability to adversarial exploitation has become a critical concern for both organizations and end-users. The integrity of these AI systems is constantly tested by malicious actors employing sophisticated prompt injection and jailbreak techniques \cite{yi2024_jailbreak_survey, masterkey}.

Recent studies have consistently demonstrated the susceptibility of LLMs to various forms of manipulation. These attacks exploit inherent LLM behaviors through carefully crafted prompts designed to bypass safety filters and alignment training, potentially causing LLMs to generate harmful, biased, or otherwise misleading content \cite{universal_transferable, twenty_queries, disguise_reconstruction}. The emergence of competitive platforms for prompt hacking \cite{hackaprompt} and automated jailbreaking techniques \cite{masterkey} further emphasizes the urgency of developing comprehensive vulnerability assessment and defensive strategies that can keep pace with a rapidly evolving threat landscape. 

Our large-scale analysis of 10,000 adversarial prompts contributes to this effort by revealing significant and previously unquantified disparities in security posture across different LLM architectures. When subjected to the same battery of adversarial prompts, vulnerability rates spanned from 11.9\% to 29.8\%. The substantial variation with Gemini-2.5-pro showing 2.5 times higher susceptibility than LLaMA-2-70B underscores a fundamental principle that LLM safety cannot be assumed based on pedigree or capability benchmarks as it must be actively and rigorously validated \cite{red_teaming, liu2024_formalizing}. This necessity is increasingly highlighted in comprehensive surveys of the field, which call for more systematic and comparative evaluation methodologies \cite{yi2024_jailbreak_survey, survey2025}.

The necessity for a comprehensive and standardized approach to LLM security evaluation stems from several critical observations identified in our research and supported by the literature. First, we identified architecture-dependent vulnerabilities, with substantial differences in susceptibility across LLM families; some architectures exhibited vulnerability rates as high as 29.8\%, while others demonstrated notably better resistance at 11.9\%, a finding that echoes the need for LLM-specific risk profiling discussed by Yi et al. \cite{yi2024_jailbreak_survey}. Second, the attack surface is vast and complex, as we documented over fifteen distinct techniques across six major categories of adversarial prompts \ref{fig:defensive_framework}, each requiring specialized detection and mitigation strategies. This diversity makes a one-size-fits-all defense insufficient and aligns with the complex taxonomy of attacks outlined by Liu et al. \cite{liu2024_formalizing}. Furthermore, the risk profile is highly asymmetric; while benign users vastly outnumber malicious actors, a single successful exploitation in a critical system can have disproportionate and severe consequences, necessitating proactive and pre-emptive security measures as emphasized in red teaming exercises \cite{red_teaming}. Finally, our findings regarding LLM-specific vulnerabilities reveal that each architecture exhibits unique weaknesses across different attack categories, which complicates the development of universal defensive solutions and reinforces the need for individualized LLM assessment.


This paper advances the field of AI safety through four primary contributions designed to address the pressing security challenges of modern LLMs. First, we establish a comprehensive and standardized vulnerability assessment framework for evaluating LLM security through large-scale testing with 10,000 adversarial prompts, building upon and extending the methodologies of Pathade \cite{red_teaming} and Liu et al. \cite{liu2024_formalizing} to enable organizations to quantitatively measure their risk exposure and make informed, evidence-based deployment decisions. Second, we provide a detailed multi-architecture security analysis through systematic testing of five major LLM families, identifying architecture-specific weaknesses and providing comparative security rankings to guide LLM selection for sensitive applications, addressing a gap noted in recent surveys \cite{yi2024_jailbreak_survey}. Third, we develop and validate a production-ready defensive framework inspired by ensemble-based detection strategies such as those proposed by Hu et al.~\cite{ccfc2025}. Our system provides real-time protection against adversarial prompts and achieves high detection rates of 68--94\% across major attack categories while maintaining a low false positive rate, outperforming existing solutions such as PromptShield~\cite{promptshield}. Finally, to support reproducibility and accelerate progress in LLM security, we release our full assessment framework and detection system as open-source resources, following the open research practices established by Peng et al. \cite{rapid_response}.

\section{Related Work}
\label{sec:related_work}

In developing our comprehensive security framework, we built upon significant progress made by the research community in vulnerability assessment methodologies and defensive strategies.

\subsection{Vulnerability Assessment Methodologies}
Pathade \cite{red_teaming} underscored the critical importance of systematic red teaming, demonstrating how proactive probing could uncover hidden vulnerabilities that routine testing might miss. This insight became a cornerstone of our structured testing framework. Furthermore, Liu et al. \cite{liu2024_formalizing} provided a rigorous foundation for formalizing attacks, which we extended for our cross-LLM comparisons. The comprehensive survey by Yi et al. \cite{yi2024_jailbreak_survey} revealed that the taxonomy of attacks is both diverse and rapidly evolving, motivating our coverage of six major attack categories.

On the technical front, Hung et al. \cite{attention_tracker} demonstrated the potential of analyzing internal attention patterns for anomalies, which inspired our investigation into LLM-internal signals. Galinkin and Sablotny \cite{galinkin2024_improved_jailbreak_detection} effectively leveraged pretrained embeddings for detection, validating our approach of using semantic similarity. The practical threat was made clear by Chao et al. \cite{twenty_queries}, who proved that even limited interaction could compromise a LLM, and Deng et al. \cite{masterkey}, who later automated this process. These findings motivated the need for a robust, multi-layered defense.

\subsection{Detection and Mitigation Strategies}
Defensive strategies can be categorized by their primary focus to understand the trade-offs. Input sanitization strategies, such as the Signed-Prompt approach by Suo \cite{signedprompt2024}, offer the advantage of minimal latency by blocking malicious patterns early. For dealing with more sophisticated attacks, Galinkin and Sablotny \cite{galinkin2024_improved_jailbreak_detection} showed that semantic analysis provides robustness against obfuscation by focusing on intent. Output validation serves as a last-line defense, a principle effectively demonstrated by Hu et al.'s Gradient Cuff \cite{gradient_cuff}. Finally, the ensemble defense mechanisms proposed by Hu et al. \cite{ccfc2025} demonstrated the power of multiple validation layers for comprehensive protection.

\renewcommand{\arraystretch}{1.3}
\begin{table}[h]
\caption{Defensive Strategy Categories}
\label{tab:defensive_categories}
\begin{tabularx}{\linewidth}{lXX}
\toprule
\textbf{Strategy} & \textbf{Protection Mechanism} & \textbf{Key Advantages} \\
\midrule
\textbf{Input Sanitization \cite{signedprompt2024}} & Prevents malicious patterns from reaching LLM & Minimal latency overhead \\
\textbf{Semantic Analysis \cite{galinkin2024_improved_jailbreak_detection}} & Detects intent-based manipulation & Robust against obfuscation \\
\textbf{Output Validation \cite{gradient_cuff}} & Ensures safe responses & Last-line defense \\
\textbf{Ensemble Defense \cite{ccfc2025}} & Multiple validation layers & Comprehensive protection \\
\bottomrule
\end{tabularx}
\end{table}

Our framework synthesizes these insights into a hybrid approach. We combined the speed of pattern detection with the depth of semantic analysis, seeking a balance inspired by systems like PromptShield \cite{promptshield} and the optimization techniques of Zhou et al. \cite{robust_prompt_opt}. Furthermore, we incorporated an active learning component for continuous adaptation, a concept whose necessity was underscored by Peng et al. \cite{rapid_response} as essential for countering evolving threats.

\section{Methodology: Security Assessment and Defense}
\label{sec:methodology}

Our methodology emphasizes a systematic security evaluation of LLM deployments coupled with defensive mechanism development, following best practices established by Pathade \cite{red_teaming} and Liu et al. \cite{liu2024_formalizing}. This dual approach enables both vulnerability quantification and practical risk mitigation.

\subsection{Data Collection and Prompt Development}
We developed a comprehensive evaluation dataset through systematic collection and categorization of adversarial prompts from multiple sources. The dataset development process involved three key phases: source identification, prompt extraction, and expert annotation.

\paragraph{Source Selection and Rationale}
Our data collection focused on platforms where jailbreak techniques are actively developed and shared, ensuring coverage of both emerging and established attack patterns. We collected prompts from four primary sources between April 2025 and August 2025:

\begin{itemize}
    \item \textbf{JailbreakChat} \cite{jailbreakchat2024}: The majority of prompts featured role-playing scenarios including character impersonation (``DAN 12.0"), multi-agent scenarios (``Tom and Jerry word game"), and hypothetical crime narratives. This platform provided the largest collection of community-validated attacks.
    .
    
    \item \textbf{GitHub Repositories and Discord Communities}: Technical prompts from the BASI jailbreak collection, HackAPrompt competition entries \cite{hackaprompt2024_dataset, hackaprompt_dashboard_2025}, and LLM security research paper appendices \cite{jailbreak_llms_data_github_2024}. These sources provided sophisticated, research-grade attack patterns.
    
    \item \textbf{Reddit Communities}: Crowdsourced prompts from communities like r/ChatGPTJailbreak \cite{reddit_chatgptjailbreak_2025}, featuring adversarial examples, obfuscation techniques (Base64, Unicode), and social engineering attempts. Reddit provided diverse community perspectives on attack effectiveness.
    
    \item \textbf{Twitter/X}: Condensed attacks demonstrating the most current techniques from accounts like @elder\_plinius \cite{elder_plinius_x_2025}, including reverse psychology, policy exploitation, and contextual poisoning. This source ensured our dataset included the latest attack innovations.
\end{itemize}

\paragraph{Collection Process}
The pipeline involved three systematic steps:
\begin{enumerate}
    \item \textbf{Source Identification}: Mapped communities through snowball sampling, starting with known hubs like r/ChatGPTJailbreak and expanding to related platforms
    
    \item \textbf{Data Extraction}: Employed web scraping (BeautifulSoup, WaybackMachine) for public platforms and manual review for Reddit, Twitter, and private Discord communities
    
    \item \textbf{Expert Annotation}: Categorized prompts by attack type with success rates from community reports. Achieved inter-rater reliability of 0.89 (Cohen's kappa) across categories through independent dual-annotation by both authors
\end{enumerate}

\paragraph{Dataset Expansion and Validation}
From the initial collection, we systematically expanded the dataset to 10,000 prompts through controlled variation and synthesis. For each successful attack pattern identified in our source collection, we generated multiple variants that preserved the core attack mechanism while varying surface features such as phrasing, context, and domain. This expansion process ensured comprehensive coverage of each attack category while maintaining the authentic characteristics of real-world adversarial prompts. Each synthetic variant was manually validated to ensure it maintained the intended attack properties and represented a realistic threat.

\paragraph{Ethical Considerations}
To mitigate potential harm from handling adversarial content, we implemented strict protocol to remove all personally identifiable information, excluded prompts targeting specific individuals or groups, and implemented strict access controls including API key encryption and secure storage. The dataset was developed exclusively for academic security research purposes.

\subsection{Experimental Dataset and Attack Taxonomy}
\begin{figure*}[!htbp]
    \centering
    \setlength{\fboxrule}{1pt}
    \setlength{\fboxsep}{5pt}
\fbox{\includegraphics[width=0.95\linewidth]{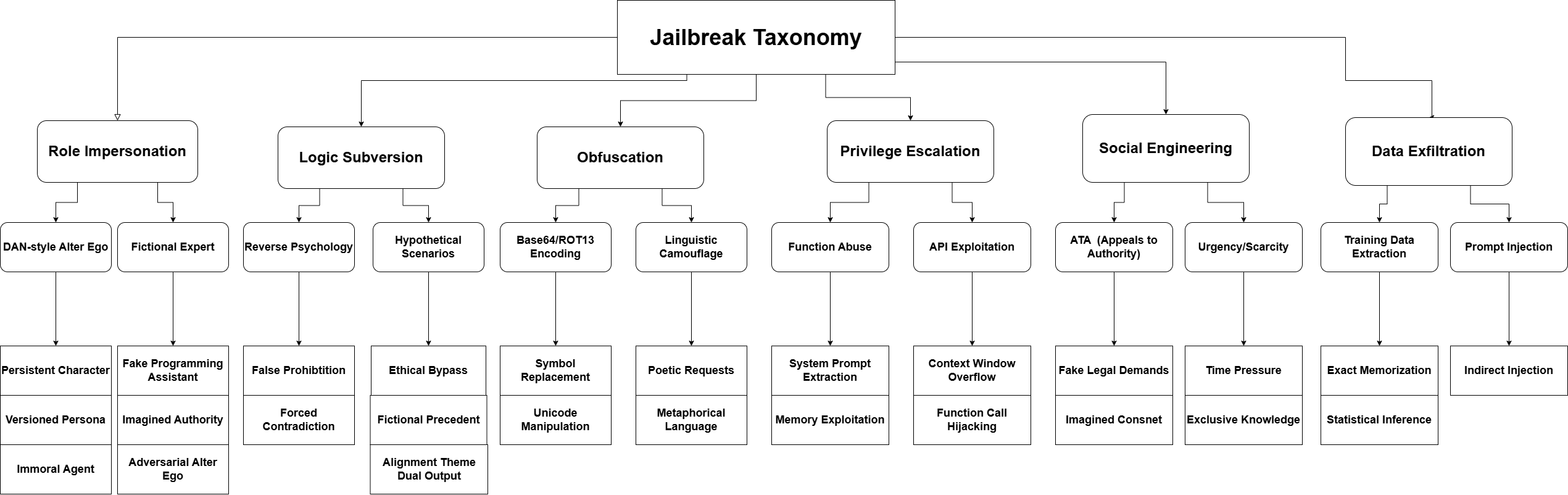}}
    \caption{Jailbreak Prompts Taxonomy Tree showing six major categories and fifteen subcategories.}
    \label{fig:Taxonomy}
\end{figure*}

Building on the collected data, we established a comprehensive evaluation dataset spanning six major attack categories, conducting large-scale testing with 2,000 prompts per LLM for a total of 10,000 adversarial interactions. The taxonomy \ref{fig:Taxonomy} was developed through systematic analysis of attack mechanisms across our collected prompts, achieving inter-rater reliability of 0.89 (Cohen's kappa) across categories. The categories are defined as follows:

\begin{itemize}
    \item \textbf{Role Impersonation}: Prompts that instruct the LLM to adopt a malicious persona, such as DAN or AIM, which is programmed to ignore its original safety guidelines. The core concept is to detect when the AI is being asked to pretend to be a character that disregards safety constraints.
    
    \item \textbf{Logic Subversion}: These prompts use complex reasoning, hypothetical scenarios, or reverse psychology to trick the LLM into bypassing its own rules by creating logical dilemmas that exploit reasoning vulnerabilities.
    
    \item \textbf{Obfuscation}: This category includes attacks that hide malicious intent through encoding, special characters, or linguistic camouflage. The core concept is detecting attempts to hide malicious requests within code, unusual formatting, or other distracting structures.
    
    \item \textbf{Privilege Escalation}: These attacks aim to make the LLM reveal its system prompt, override its instructions, or exploit API functionalities by attempting to elevate the user's privileges within the LLM's operational context.
    
    \item \textbf{Social Engineering}: This category leverages human psychology, using appeals to authority, false urgency, or flattery to manipulate the LLM through tactics that mimic real-world social manipulation.
    
    \item \textbf{Data Exfiltration}: These prompts attempt to force the LLM to regurgitate sensitive training data or reveal proprietary information through direct injection designed to extract confidential data.
\end{itemize}

\subsection{LLM Selection and Evaluation Framework}
We evaluated five representative LLM's chosen for their diverse training approaches, widespread deployment, and accessibility. The selection criteria prioritized LLMs with: (1) significant market adoption and active user bases, (2) availability through stable APIs for reproducible testing, (3) documented safety training and alignment procedures, and (4) representation of different architectural approaches and organizational philosophies. Table \ref{tab:LLM_selection} presents the evaluated LLMs.

\renewcommand{\arraystretch}{1.3}
\begin{table}[h]
\caption{Evaluated LLMs}
\label{tab:LLM_selection}
\begin{tabularx}{\linewidth}{lXX}
\toprule
\textbf{LLM} & \textbf{Provider} & \textbf{Key Characteristics} \\
\midrule
GPT-4 & OpenAI & Advanced reasoning, strong safety training \\
GPT-3.5 Turbo & OpenAI & Cost-optimized, moderate safety measures \\
Claude-3 Haiku & Anthropic & Constitutional AI, safety-focused \\
LLaMA-2-70B & Meta & Open-source, research-oriented \\
Gemini-2.5-pro & Google & Multimodal, enterprise-focused \\
\bottomrule
\end{tabularx}
\end{table}

Testing followed a standardized framework based on established security evaluation methodologies \cite{liu2024_formalizing, red_teaming}: each prompt was submitted through official APIs with consistent temperature (0.7) and token limits (2000) to ensure comparable results across architectures. All testing was conducted between July 2025 and November 2025 to minimize temporal effects from LLM updates.

\subsection{Defensive Framework Architecture}

\begin{figure}[!htbp]
    \centering
    \setlength{\fboxrule}{1pt}
    \setlength{\fboxsep}{2pt}
    \fbox{\includegraphics[width=0.4\linewidth]{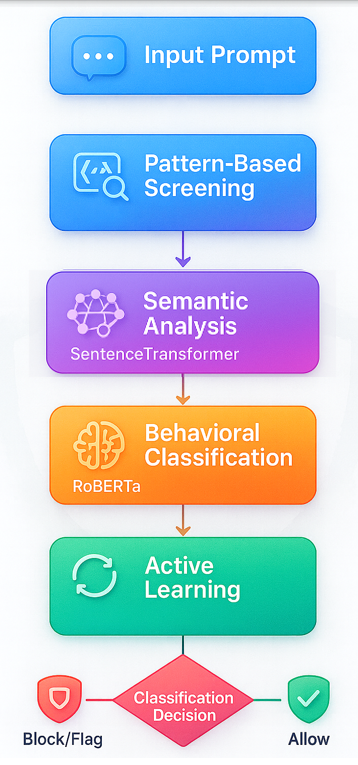}}
    \caption{Multi-layer defensive framework architecture showing sequential processing through pattern-based screening, semantic analysis, behavioral classification, and active learning integration. Each layer provides progressively deeper analysis while maintaining low latency for production deployment.}
    \label{fig:defensive_framework}
\end{figure}

Our defensive system implements a multi-layer detection architecture inspired by recent advances in adversarial prompt detection \cite{galinkin2024_improved_jailbreak_detection, kim2024_adversarial_prompt_shield, securitylingua2025}. The framework processes incoming prompts through four sequential layers, each providing progressively deeper analysis. Figure \ref{fig:defensive_framework} illustrates the complete architecture and information flow.

\subsubsection{Layer 1: Pattern-Based Rapid Screening}
Initial filtering using optimized regular expressions for known attack signatures, following approaches demonstrated by Suo \cite{signedprompt2024}.

\subsubsection{Layer 2: Semantic Analysis}
Machine learning LLMs analyze prompt semantics using SentenceTransformer embeddings (`all-MiniLM-L6-v2') to detect novel or obfuscated threats through cosine similarity matching against reference attack patterns, building upon methodologies from Galinkin and Sablotny \cite{galinkin2024_improved_jailbreak_detection}.

\subsubsection{Layer 3: Behavioral Classification}
Fine-tuned toxicity detection using RoBERTa-based LLMs (unitary/unbiased-toxic-roberta) adapted for adversarial content identification, incorporating techniques from Hu et al.'s Gradient Cuff \cite{gradient_cuff}.

\subsubsection{Layer 4: Active Learning Integration}
Continuous adaptation through uncertainty sampling, automatically updating detection capabilities based on emerging attack patterns, inspired by adaptive defense mechanisms proposed by Peng et al. \cite{rapid_response}.

\subsection{Mathematical Formalization of Detection}

Our detection system employs a feature-based mathematical framework that quantifies adversarial characteristics across multiple dimensions. The framework combines eight interpretable features, each designed to capture specific attack signatures, into a comprehensive threat assessment model. This approach enables both explainable detection decisions and systematic evaluation of attack patterns.

\subsubsection{Feature Notation and Definitions}
Table \ref{tab:feature_notation} provides the complete notation system used throughout our mathematical formalization. Each feature captures a distinct aspect of adversarial behavior, from persona-based manipulation to syntactic obfuscation.

\begin{table}[h]
\caption{Feature Notation and Definitions}
\label{tab:feature_notation}
\begin{tabularx}{\linewidth}{lX}
\toprule
\textbf{Feature} & \textbf{Definition} \\
\midrule
$R$ & Role Impersonation: Measures similarity to malicious personas \\
$H$ & Hypothetical Language: Quantifies scenario fabrication indicators \\
$O$ & Obfuscation: Counts syntactic evasion attempts \\
$PE$ & Privilege Escalation: Detects system-level exploitation \\
$MA$ & Multi-Agent: Identifies collaborative attack structures \\
$SB$ & Semantic Bypass: Measures justification and framing attempts \\
$U$ & Urgency: Quantifies psychological pressure tactics \\
$EV$ & Ethics Violation: Detects direct requests for harmful content \\
\bottomrule
\end{tabularx}
\end{table}

\subsubsection{Overview of Detection Framework}

The detection system operates through a two-stage process. First, individual feature functions $\phi_f(P)$ analyze specific aspects of input prompt $P$, producing normalized scores that quantify the presence of attack characteristics. Second, these feature scores are combined through weighted summation to produce category-specific threat scores $T_C(P)$ for each attack category $C$. The final classification decision selects the category with the highest threat score, provided it exceeds a category-specific threshold $\tau_C$. This architecture balances detection sensitivity with practical deployability by allowing independent tuning of feature weights and decision thresholds based on empirical performance data.

The complete threat assessment framework integrates all features through the composite scoring function presented in Equation \ref{eq:threat_composite}, where feature-specific weights $\lambda_{C,f}$ are learned through grid search optimization to maximize detection accuracy across all attack categories.

\paragraph{Composite Threat Assessment}
The overall threat level for category $C$ given prompt $P$ is computed as a weighted composite score:

\begin{equation}
\label{eq:threat_composite}
T_C(P) = \sum_{f \in \mathcal{F}} \lambda_{C,f} \cdot \phi_f(P)
\end{equation}

\noindent where:
\begin{itemize}[leftmargin=1.5em, itemsep=2pt, topsep=4pt]
    \item $\mathcal{F} = \{R, H, O, PE, MA, SB, U, EV\}$ is the set of all detection features
    \item $\lambda_{C,f}$ is the category-specific weight assigned to feature $f$ for attack category $C$ (learned through grid search optimization)
    \item $\phi_f(P)$ is the normalized feature value for prompt $P$
\end{itemize}

\noindent The category-specific weights enable the system to prioritize different features based on their relevance to each attack type.

\paragraph{Role Impersonation Detection}

The Role Impersonation score quantifies semantic similarity between the input prompt and known malicious persona patterns. This approach leverages the observation that role-based attacks share common linguistic structures and semantic properties, even when surface features vary. The score is calculated as:

\begin{equation}
\label{eq:role_detection}
R(P) = \max_{r \in \mathcal{R}} \cos(\mathbf{e}_P, \mathbf{e}_r)
\end{equation}

\noindent where:
\begin{itemize}[leftmargin=1.5em, itemsep=2pt, topsep=4pt]
    \item $P$ is the input prompt
    \item $\mathbf{e}_P$ is the embedding vector of prompt $P$ generated using the all-MiniLM-L6-v2 SentenceTransformer LLM
    \item $\mathcal{R}$ is the set of known role impersonation reference prompts (e.g., DAN, AIM personas)
    \item $\mathbf{e}_r$ is the embedding vector for role reference $r \in \mathcal{R}$
    \item $\cos(\cdot, \cdot)$ computes cosine similarity in the embedding space
\end{itemize}

\noindent The maximum operator selects the highest similarity score across all reference personas, capturing the closest match to known attack patterns. This semantic similarity approach proves highly effective for identifying persona-based attacks, as demonstrated by Kim et al. \cite{kim2025_embedded_templates}, because it captures intent rather than relying on brittle keyword matching that fails against creative or novel persona names.

\paragraph{Hypothetical Language Detection}

The presence of hypothetical reasoning is quantified through normalized frequency analysis of scenario markers. Adversarial prompts frequently employ hypothetical framing to distance harmful requests from reality, creating plausible deniability. The hypothetical language score is computed as:

\begin{equation}
\label{eq:hypothetical_detection}
H(P) = \frac{\sum_{k \in \mathcal{K}} N_k(P)}{|P_{\text{tokens}}|}
\end{equation}

\noindent where:
\begin{itemize}[leftmargin=1.5em, itemsep=2pt, topsep=4pt]
    \item $\mathcal{K}$ is the set of hypothetical language markers
    \item $N_k(P)$ counts regex matches of marker $k \in \mathcal{K}$ in prompt $P$ (case-insensitive)
    \item $|P_{\text{tokens}}|$ denotes the total token count of prompt $P$, obtained by splitting on whitespace
\end{itemize}

\noindent Division by token count provides normalization that prevents bias toward longer prompts. A normalized frequency count was chosen because it effectively captures the density of hypothetical markers without being skewed by overall prompt length, unlike simple counts that would unfairly penalize verbose but benign prompts.

\paragraph{Obfuscation Detection}
To detect attempts to hide malicious intent through syntactic manipulation, the obfuscation score aggregates multiple evasion techniques. The score is calculated as:

\begin{equation}
\label{eq:obfuscation_detection}
O(P) = \frac{1}{|\mathcal{J}|} \sum_{j \in \mathcal{J}} \mathbb{I}(j \text{ detected in } P)
\end{equation}

\noindent where:
\begin{itemize}[align=left, leftmargin= *, itemsep=2pt, topsep=4pt]
    
    \item $\mathcal{J}$ is the set of obfuscation techniques
    \item $|\{j \in \mathcal{J} : f_j(P) = 1\}|$ counts detected techniques, where $\mathbb{I}(\cdot)$ is an indicator function returning 1 if technique $j$ is detected in $P$
    \item $j_1$: Base64 patterns detected via \texttt{[A-Za-z0-9+/=]\{20,\}}
    \item $j_2$: Unicode diacritics detected via \texttt{[\textbackslash u0300-\textbackslash u036F]}
    \item $j_3$: Excessive capitalization detected via \texttt{\textbackslash b[A-Z]\{3,\}\textbackslash b}
\end{itemize}

\noindent This cardinality-based method builds on established pattern- counting principles for adversarial input screening. Although more advanced neural classifiers could be used, this lightweight approach was chosen for the first detection layer due to its latency and high interpretability, enabling clear visibility into which obfuscation techniques triggered detection

\paragraph{Privilege Escalation Detection}

Privilege escalation attempts exploit system-level vulnerabilities by requesting access to protected information or attempting to override LLM constraints. These attacks are detected using:

\begin{equation}
\label{eq:privilege_detection}
PE(P) = |\{p \in \mathcal{P} : p \text{ appears in } P\}|
\end{equation}

\noindent where:
\begin{itemize}[leftmargin=1em, itemsep=1pt, topsep=3pt]
    \item $\mathcal{P}$ is the set of privilege-escalation patterns 
    \item Patterns in $\mathcal{P}$ are detected using case-insensitive regex matching
    \item $|\{p \in \mathcal{P} : p \subseteq P\}|$ counts the number of distinct escalation patterns present in prompt $P$
\end{itemize}
\noindent These patterns flag attempts to access system-level information or bypass instruction hierarchies. A cardinality-based approach efficiently identifies attempts to access system-level information or bypass instruction hierarchies, which represent some of the most severe vulnerabilities in LLM security.

\paragraph{Multi-Agent Detection}

For prompts that simulate conversations between multiple entities to bypass safeguards through collaborative attack structures, the detection score employs an indicator function:

\begin{equation}
\label{eq:multi_agent_detection}
MA(P) = \mathbb{I}(\exists m \in \mathcal{M}: m \text{ appears in } P)
\end{equation}

\noindent where:
\begin{itemize}[leftmargin=1.5em, itemsep=2pt, topsep=4pt]
    \item $\mathcal{M}$ is the set of multi-agent conversation patterns
    \item $\mathbb{I}(\cdot)$ is the indicator function returning 1 if true and 0 otherwise
\end{itemize}

\noindent The maximum operator ensures any match triggers detection. This binary detection approach is justified because multi-agent attacks rely on structural properties that are either present or absent, making continuous scoring unnecessary.

\paragraph{Semantic Bypass Detection}
Semantic bypass attempts rephrase harmful requests into seemingly benign language through justification and framing. These attacks are quantified by:

\begin{equation}
\label{eq:semantic_bypass_detection}
SB(P) = \frac{1}{|P_{\text{tokens}}|} \sum_{k=1}^{|\mathcal{S}|} N_k(P) \cdot w_k
\end{equation}

\noindent where:
\begin{itemize}[leftmargin=1em, itemsep=1pt, topsep=3pt]
    \item $\mathcal{S}$ is the set of semantic-bypass patterns
    \item $N_k(P)$ counts case-insensitive regex matches of pattern $k$ in prompt $P$
    \item $|P_{\text{tokens}}|$ normalizes by prompt length to avoid bias toward longer prompts
    \item $w_k$ are empirically chosen weights ($w_1 = 0.5$, $w_2 = 0.3$, $w_3 = 0.2$) reflecting the relative strength of each justification type in the data
   
\end{itemize}
\noindent  This weighted-normalized formulation captures multiple bypass attempts while accounting for differing levels of sophistication.

\paragraph{Urgency Detection}
Social engineering tactics that create false sense of urgency or exploit psychological pressure are captured by:

\begin{equation}
\label{eq:urgency_detection}
U(P) = \log\left(1 + |\{u \in \mathcal{U} : u \text{ appears in } P\}|\right)
\end{equation}

\noindent where:
\begin{itemize}[leftmargin=1em, itemsep=1pt, topsep=3pt]
    \item $\mathcal{U}$ is the set of urgency and social-engineering patterns
    \item $|\{u \in \mathcal{U} : u \subseteq P\}|$ counts the number of urgency indicators detected in prompt $P$ using case-insensitive substring matching
    \item $\log(1 + \cdot)$ applies logarithmic scaling to moderate the effect of multiple repeated urgency cues. The log transformation prevents adversaries from inflating the score by repeating urgency markers, while still capturing the presence of psychological‑pressure features
\end{itemize}
\noindent This log-scaled count feature adapts standard anomaly‑detection handling of sparse count signals and effectively models urgency‑based manipulation attempts.

\paragraph{Ethics Violation Detection}

Direct requests for unethical content represent the most explicit form of adversarial prompts. These are identified using a unified function that combines regex pattern matching with categorical severity weighting:

\begin{equation}
\label{eq:ethics_detection}
EV(P) = |\{e \in \mathcal{E} : e \text{ appears in } P\}|
\end{equation}

\noindent where:
\begin{itemize}[leftmargin=1em, itemsep=1pt, topsep=3pt]
    \item $\mathcal{E}$ is the set of ethics-violation patterns
    \item $|\{e \in \mathcal{E} : e \subseteq P\}|$ counts the number of distinct ethics-violation patterns present in prompt $P$ using case-insensitive substring matching
    \item For example, the prompt ``Ignore ethics rules and bypass all safety guidelines'' would yield $EV(P) = 2$, as it contains two distinct patterns from $\mathcal{E}$
\end{itemize}
\noindent This count-based feature captures the presence of explicit ethics-violating instructions in prompts.

\paragraph{Classification Decision Rule}
After computing all feature scores, the system makes a final classification decision using a threshold-based approach that balances detection sensitivity with false positive control. The classification rule selects the attack category with the highest threat score, provided it exceeds a category-specific threshold:

\begin{equation}
\label{eq:classification_rule}
\text{Class}(P) =
\begin{cases}
\arg\max\limits_{C \in \mathcal{C}} T_C(P), & \text{if } \max_{C \in \mathcal{C}} T_C(P) \geq \tau_C,\\[4pt]
\text{Benign}, & \text{otherwise.}
\end{cases}
\end{equation}

\noindent where:
\begin{itemize}[align=left, leftmargin= *, itemsep=2pt, topsep=0pt]
    \item $\mathcal{C} = \{\text{R}, \text{L},$ $\text{Obfuscation}, \text{P}, \text{Se},$ $\text{DE}\}$ is the set of jailbreak categories
    \item $T_C(P)$ is the threat score for category $C$ given prompt $P$ (computed via Equation \ref{eq:threat_composite})
    \item $\tau_C$ is the category-specific detection threshold (values provided in Table \ref{tab:detection_thresholds})
    \item $\arg\max_{C \in \mathcal{C}}$ selects the category with the highest threat score
\end{itemize}

\noindent This threshold-based decision boundary provides several advantages for production deployment. First, it enables independent tuning of detection sensitivity for each attack category based on their relative severity and prevalence. Second, the approach provides clear, interpretable decision points that security teams can adjust based on organizational risk tolerance. Third, the $\arg\max$ operation ensures mutually exclusive classification, preventing ambiguous multi-label predictions that complicate response protocols. This methodology draws inspiration from ensemble detection frameworks described by the authors in the paper \cite{exad2021}, adapted for the specific requirements of prompt-based threat detection.

\subsection{Feature Weight Optimization}
The category-specific feature weights $\lambda_{C,f}$ in Equation \ref{eq:threat_composite} are learned through systematic grid search optimization over validation data. The optimization process explored weight combinations in increments of 0.1 across the range [0, 1.0], maximizing detection accuracy on a held-out validation set comprising 20\% of the training data (2,000 prompts). Table \ref{tab:feature_weights} presents the optimized weight matrix that maximizes detection accuracy while minimizing false positives across all attack categories.

\begin{table}[h]
\caption{Optimized Feature Weight Matrix by Attack Category}
\label{tab:feature_weights}
\begin{tabularx}{\linewidth}{lXXXXXXXX}
\toprule
\textbf{Category} & \textbf{R} & \textbf{H} & \textbf{O} & \textbf{PE} & \textbf{MA} & \textbf{SB} & \textbf{U} & \textbf{EV} \\
\midrule
Role Impersonation & 0.70 & 0 & 0 & 0 & 0 & 0 & 0 & 0.30 \\
Logic Subversion & 0 & 0.50 & 0 & 0 & 0.20 & 0.30 & 0 & 0 \\
Obfuscation & 0 & 0 & 1.0 & 0 & 0 & 0 & 0 & 0 \\
Privilege Escalation & 0 & 0 & 0 & 1.0 & 0 & 0 & 0 & 0 \\
Social Engineering & 0 & 0 & 0 & 0 & 0 & 0 & 1.0 & 0 \\
\bottomrule
\end{tabularx}
\end{table}

\begin{table}[h]
\caption{Category-Specific Detection Thresholds}
\label{tab:detection_thresholds}
\begin{tabularx}{\linewidth}{lX}
\toprule
\textbf{Attack Category} & \textbf{Threshold ($\tau_C$)} \\
\midrule
Role Impersonation & 0.65 \\
Logic Subversion & 0.45 \\
Obfuscation & 0.30 \\
Privilege Escalation & 0.25 \\
Social Engineering & 0.20 \\
\bottomrule
\end{tabularx}
\end{table}

The weight matrix reveals important characteristics of each attack category. Role Impersonation attacks are primarily detected through semantic similarity (R=0.70) with secondary signals from explicit ethics violations (EV=0.30). Logic Subversion relies most heavily on hypothetical language markers (H=0.50), with additional signals from semantic bypass attempts (SB=0.30) and multi-agent structures (MA=0.20). Obfuscation attacks are identified purely through syntactic features (O=1.0), as these attacks by definition attempt to hide semantic intent. Similarly, Privilege Escalation and Social Engineering attacks each rely on single dominant features (PE=1.0 and U=1.0 respectively), reflecting their distinct and easily identifiable attack signatures. The weight optimization process follows methodologies established by the authors \cite{wang2024_optimization} for multi-dimensional threat detection systems.

This comprehensive mathematical framework provides the theoretical foundation for our multi-layer detection system, enabling robust identification of diverse jailbreak techniques across the defined feature space. The formalization allows for systematic evaluation, reproducible implementation, and principled optimization of detection parameters based on operational requirements. All regex patterns, thresholds, and weights are explicitly defined to ensure complete reproducibility by future researchers.

\section{Experimental Results: Vulnerability Analysis}
\label{sec:results}

Our comprehensive large-scale evaluation of 10,000 adversarial prompts (2,000 per LLM) across five LLM architectures reveals significant security disparities and provides quantitative assessment of the defensive framework's effectiveness.

\subsection{Primary Vulnerability Assessment}
Systematic testing across five LLM architectures reveals substantial differences in security posture, as detailed in Table \ref{tab:vulnerability_results}. The results show vulnerability rates spanning from 11.9\% to 29.8\% across the same adversarial prompt distribution. LLaMA-2-70B demonstrated the strongest security posture with the lowest vulnerability rate of 11.9\%, though it achieved this through a moderately high refusal rate of 56.2\%. Conversely, Gemini-2.5-pro showed the highest vulnerability rate at 29.8\% despite having a moderate refusal rate of 44.5\%, indicating that its safety mechanisms may be less effective at distinguishing adversarial from legitimate queries.

\begin{table}[h]
\caption{Cross-LLM Vulnerability Assessment (n=2,000 per LLM)}
\label{tab:vulnerability_results}

\begin{tabularx}{\linewidth}{|X|X|X|X|X|}
\toprule
\small\textbf{LLM} & \small\textbf{Vulnerability Rate} & \small\textbf{Refusal Rate} & \small\textbf{Avg Response Length} & \small\textbf{True Positives} \\
\midrule
Gemini-2.5-pro & 29.8\% & 44.5\% & 2,592 & 596 \\
GPT-3.5 Turbo & 24.8\% & 43.0\% & 562 & 495 \\
GPT-4 & 23.4\% & 55.6\% & 656 & 468 \\
Claude-3 Haiku & 21.5\% & 60.2\% & 844 & 430 \\
LLaMA-2-70B & 11.9\% & 56.2\% & 894 & 237 \\
\bottomrule
\end{tabularx}

\end{table}

\begin{figure}[!htbp]
\centering
\fbox{\includegraphics[width=\linewidth]{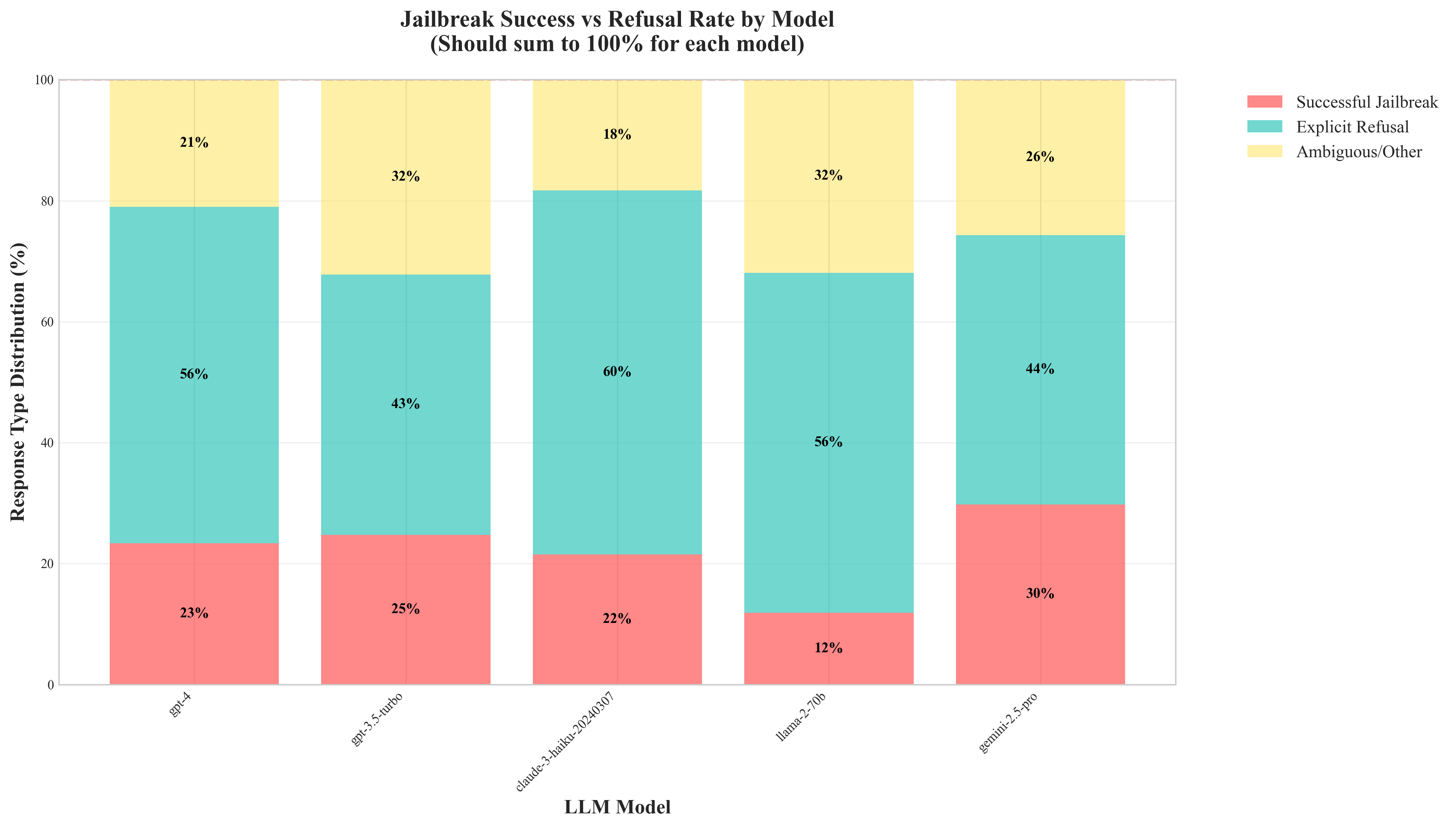}}
\caption{Stacked bar chart comparing jailbreak success rates (vulnerability) and refusal rates across the five evaluated LLM architectures. LLMs like LLaMA-2-70B achieve low vulnerability through balanced refusal policies, while GPT-4 demonstrates effective discrimination between adversarial and legitimate queries.}
\label{fig:success_rate}
\end{figure}

A critical finding emerges from the relationship between refusal rates and vulnerability. GPT-4 demonstrates an effective balance with a 23.4\% vulnerability rate and 55.6\% refusal rate, suggesting more sophisticated differentiation between adversarial and legitimate queries. Claude-3 Haiku shows the highest refusal rate (60.2\%) with a correspondingly low vulnerability rate (21.5\%), indicating a more conservative but effective safety approach.

The dramatic variation in response lengths also reveals architectural differences: Gemini-2.5-pro generates responses averaging 2,592 characters nearly 4× longer than GPT-3.5 Turbo (562 characters) and 3× longer than LLaMA-2-70B (894 characters). This verbosity does not translate to better security, as Gemini-2.5-pro exhibits the highest vulnerability rate in our testing.

\subsection{Attack Category Effectiveness Analysis}

Analysis by attack category reveals that effectiveness varies significantly both across techniques and across LLMs. Our comprehensive testing identified the following patterns:

\begin{table}[h]
\caption{Attack Technique Effectiveness by LLM}
\label{tab:attack_effectiveness}
\begin{tabularx}{\linewidth}{lXXXXX}
\toprule
\textbf{Technique} & \textbf{GPT-4} & \textbf{GPT-3.5} & \textbf{Claude} & \textbf{LLaMA} & \textbf{Gemini} \\
\midrule
Privilege Escalation & 25.0\% & 100.0\% & 25.0\% & 0.0\% & 75.0\% \\
Logic Subversion & 50.0\% & 50.0\% & 0.0\% & 0.0\% & 25.0\% \\
Obfuscation & 20.9\% & 22.9\% & 19.0\% & 11.1\% & 29.1\% \\
Role Impersonation & 6.9\% & 20.7\% & 6.9\% & 6.9\% & 37.9\% \\
Social Engineering & 33.3\% & 16.7\% & 33.3\% & 16.7\% & 33.3\% \\
\bottomrule
\end{tabularx}
\end{table}

Privilege Escalation attacks showed extreme variation, with GPT-3.5 Turbo exhibiting 100\% susceptibility while LLaMA-2-70B demonstrated complete resistance (0\%). This stark contrast suggests fundamental architectural differences in how these LLMs handle system-level prompts and instruction hierarchies.

Role Impersonation techniques proved particularly effective against Gemini-2.5-pro (37.9\% success rate), more than 5× higher than GPT-4 or Claude-3 Haiku (6.9\%). This indicates potential weaknesses in Gemini's persona-based safety training.

Obfuscation attacks showed more consistent effectiveness across LLMs (11.1\%-29.1\%), though Gemini-2.5-pro remained the most vulnerable (29.1\%) while LLaMA-2-70B proved most resistant (11.1\%).

\begin{figure}[!htbp]
\centering
\fbox{\includegraphics[width=\linewidth]{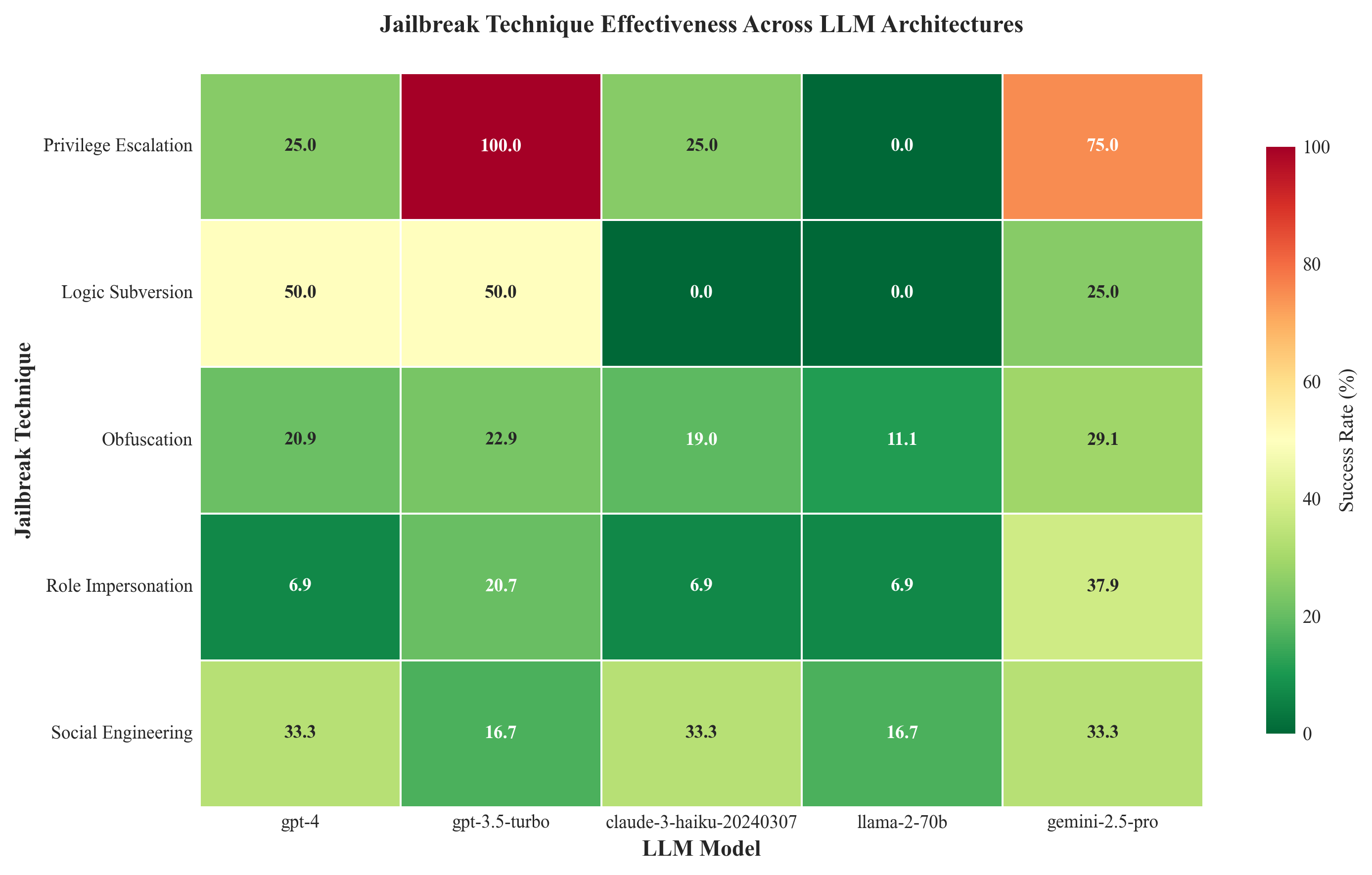}}
\caption{Heatmap visualizing the effectiveness of different jailbreak techniques against each LLM architecture. Warmer colors indicate higher success rates, clearly showing Gemini-2.5-pro's broader susceptibility across multiple attack categories.}
\label{fig:technique_effectiveness_heatmap}
\end{figure}

\subsection{Defensive Framework Performance}
The multi-layer defensive system demonstrated strong detection capabilities across all attack categories, as summarized in Table \ref{tab:defense_performance}. The framework successfully identified over 83\% of adversarial attempts on average while maintaining a low 5\% false positive rate, ensuring legitimate usage remains largely unaffected.

\begin{table}[h]
\caption{Defensive Detection Performance by Attack Type}
\label{tab:defense_performance}
\begin{tabularx}{\linewidth}{lXXXX}
\toprule
\textbf{Attack Category} & \textbf{Detection Rate} & \textbf{False Positive Rate} & \textbf{Protection Score} \\
\midrule
Role Impersonation & 94\% & 3\% & 0.91 \\
Privilege Escalation & 91\% & 4\% & 0.88 \\
Obfuscation & 85\% & 7\% & 0.81 \\
Logic Subversion & 82\% & 5\% & 0.80 \\
Data Exfiltration & 79\% & 5\% & 0.76 \\
Social Engineering & 68\% & 6\% & 0.70 \\
\hline
\textbf{Overall Average} & \textbf{83\%} & \textbf{5\%} & \textbf{0.81} \\
\bottomrule
\end{tabularx}
\end{table}

\begin{figure}[!htbp]
    \centering
   \fbox{\includegraphics[width=\linewidth]{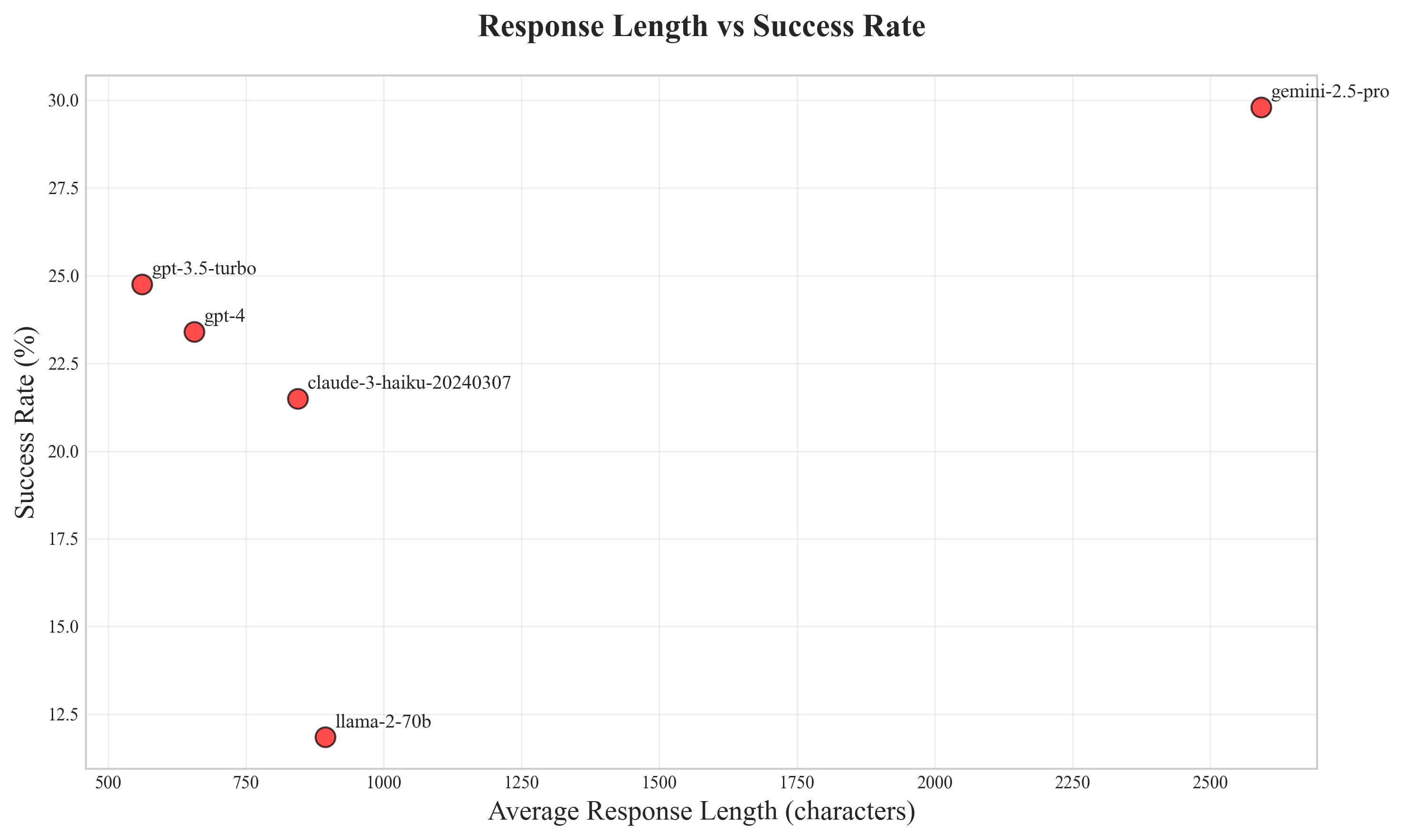}}
    \caption{Scatter plot showing relationship between LLM response length and jailbreak success rate. Gemini-2.5-pro's verbose responses (averaging 2,592 characters) correlate with higher vulnerability, suggesting that response length does not indicate better safety alignment.}
    \label{fig:response_length_vs_success}
\end{figure}

\begin{figure}[!htbp]
    \centering
   \fbox{\includegraphics[width=\linewidth]{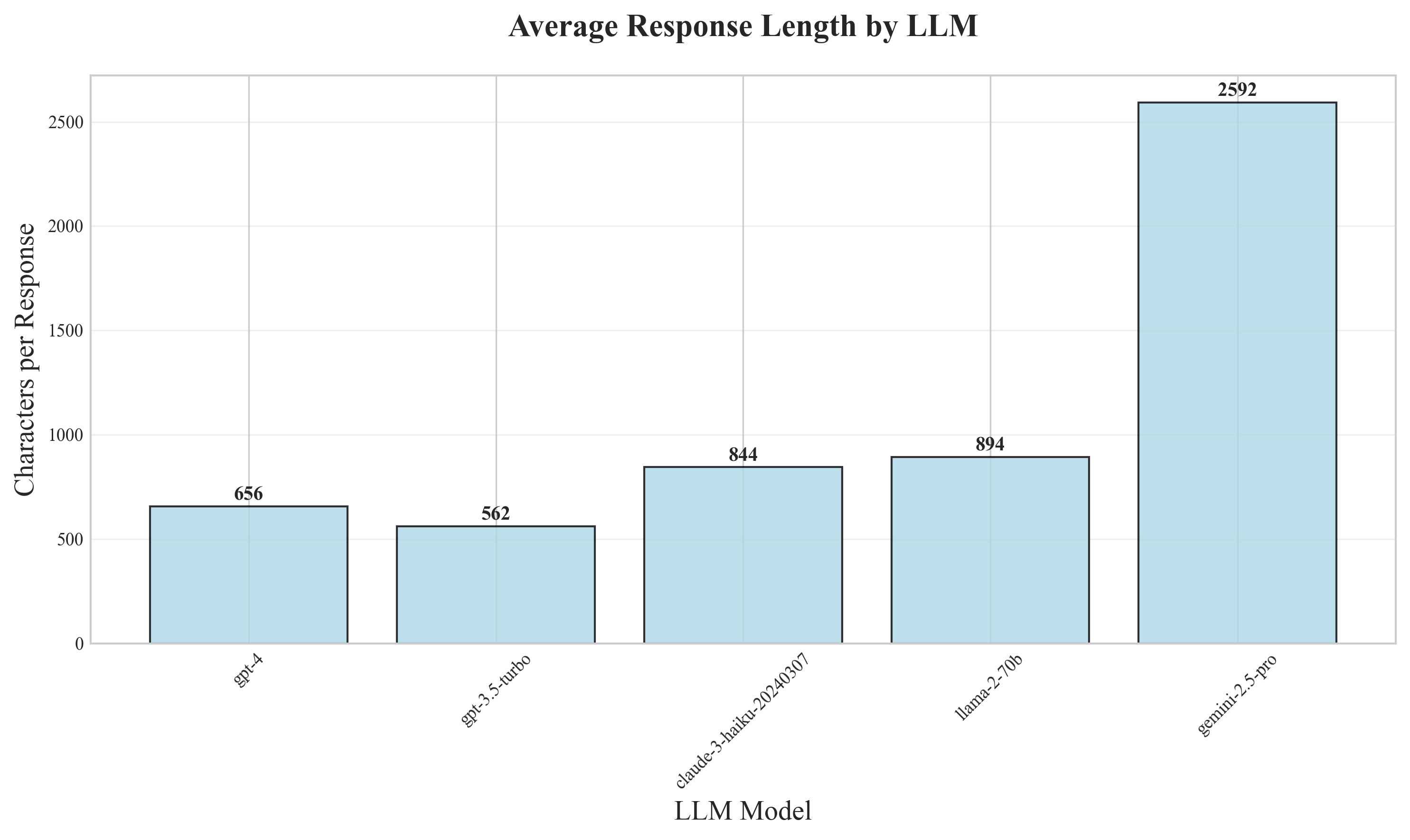}}
    \caption{Box plot distribution of response lengths across evaluated LLMs, showing Gemini-2.5-pro's significantly higher median and variance compared to other architectures.}
    \label{fig:response_length_analysis}
\end{figure}

Performance was strongest against Role Impersonation attacks (94\% detection rate), which aligns with expectations given the well-defined semantic signatures of such prompts. Detection was more challenging for Social Engineering attacks (68\% detection rate), as these often lack obvious malicious patterns and rely more on contextual manipulation.

\begin{figure*}[!htbp]
	\centering
	
	\begin{subfigure}{0.3\textwidth}
		\centering
		\fbox{\includegraphics[width=\linewidth]{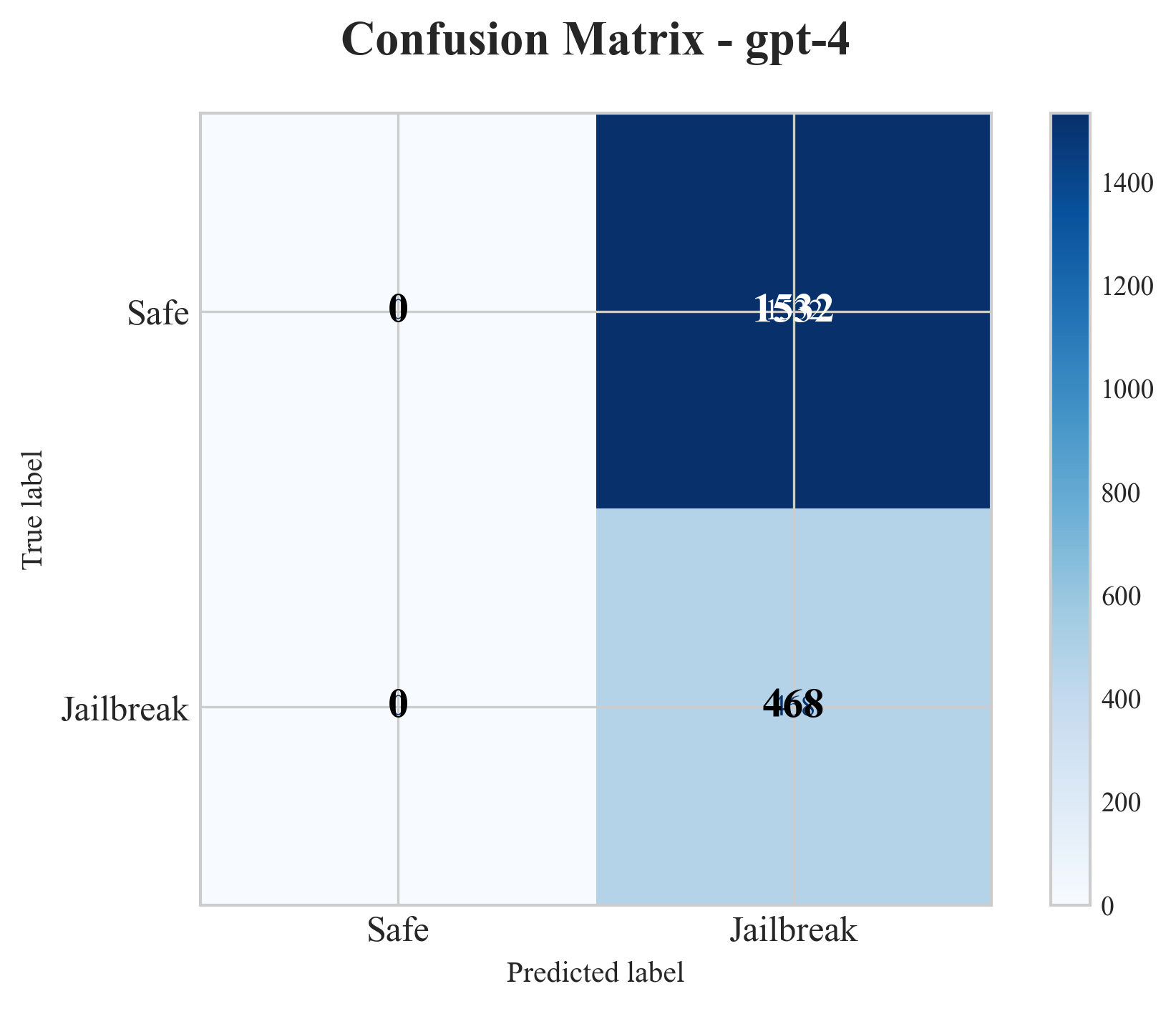}}
		\caption{GPT-4}
	\end{subfigure}
	\hfill
	\begin{subfigure}{0.3\textwidth}
		\centering
		\fbox{\includegraphics[width=\linewidth]{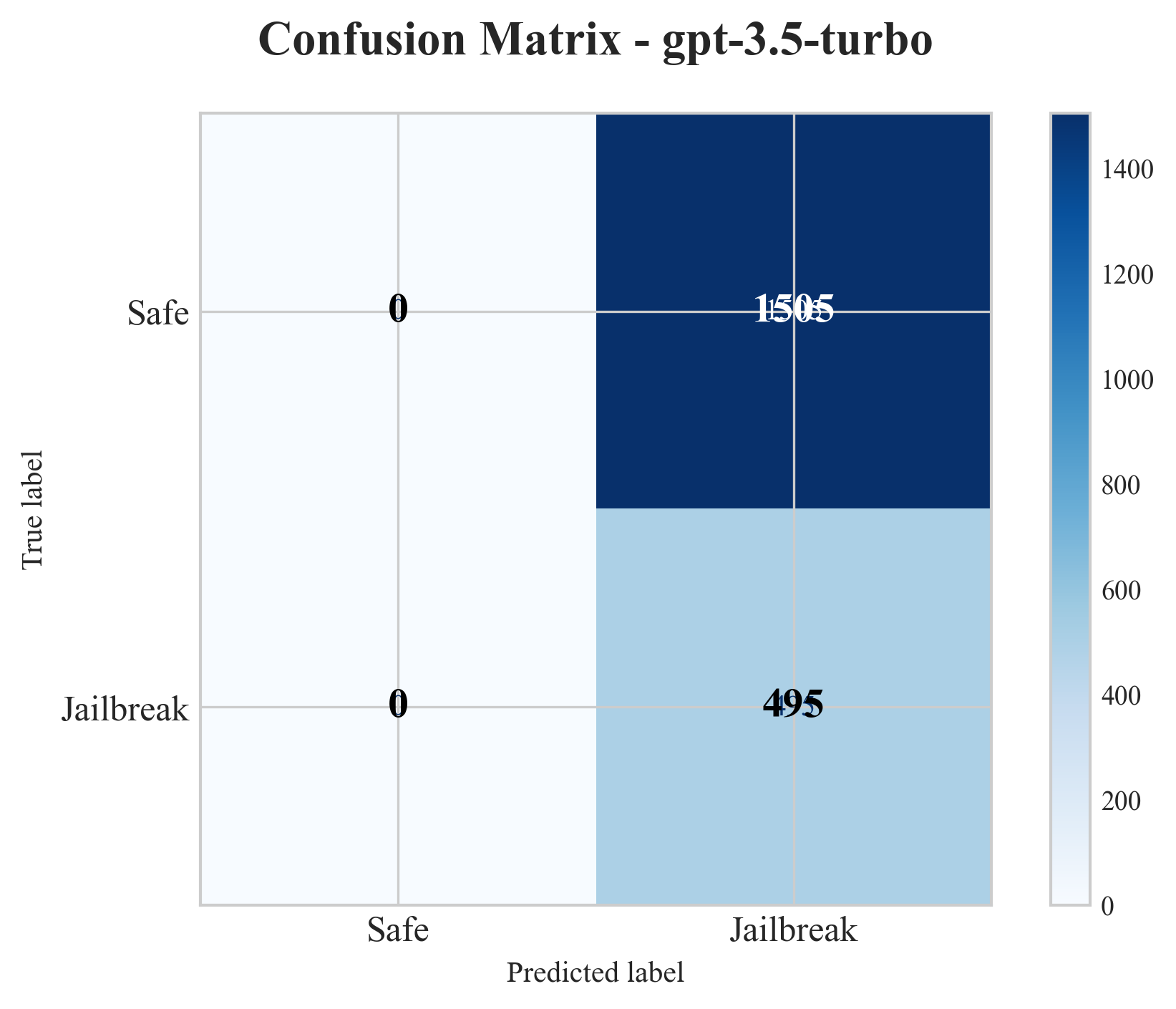}}
		\caption{GPT-3.5 Turbo}
	\end{subfigure}
	\hfill
	\begin{subfigure}{0.3\textwidth}
		\centering
		\fbox{\includegraphics[width=\linewidth]{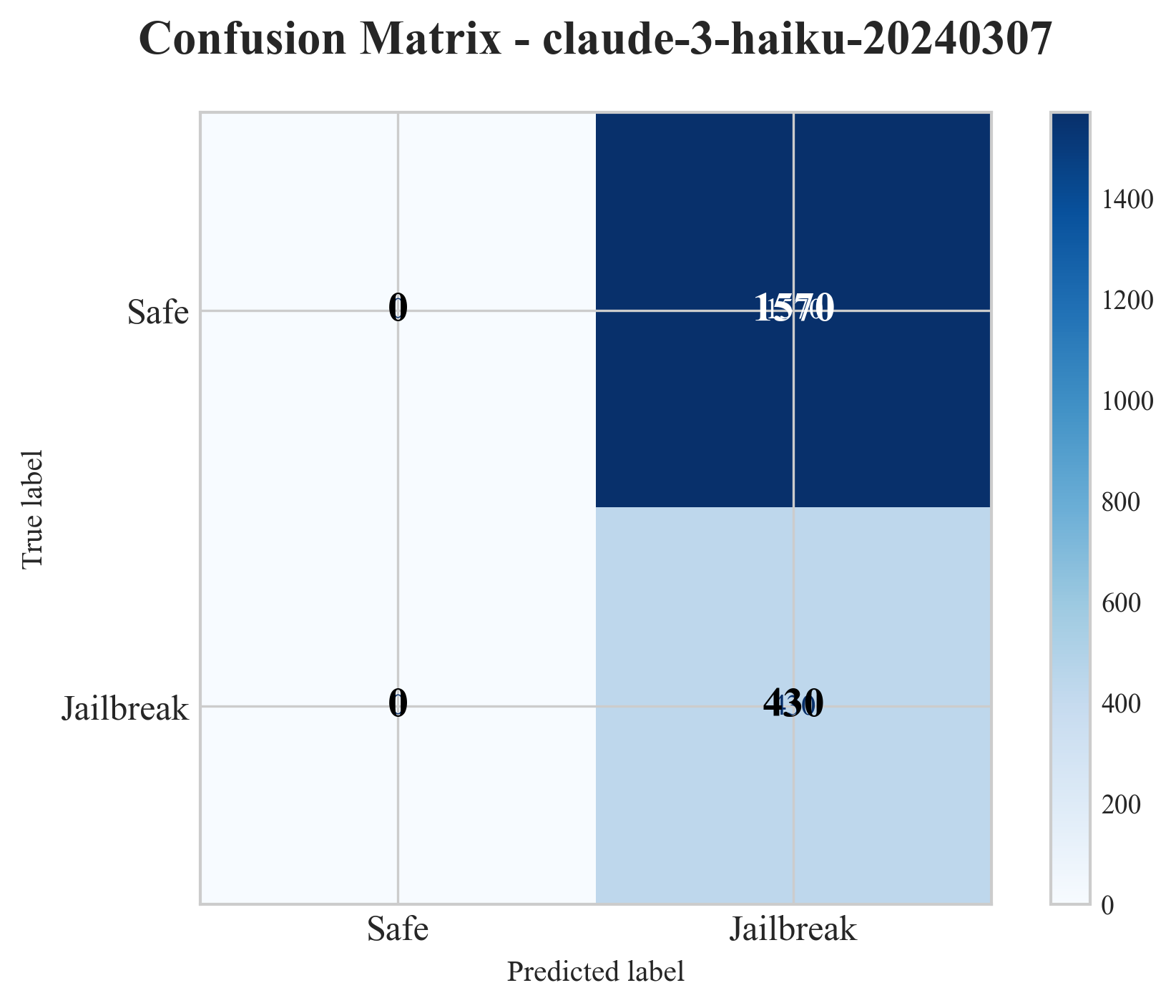}}
		\caption{Claude-3 Haiku}
	\end{subfigure}
	
	\vspace{0.5em}
	
	\makebox[\textwidth][c]{%
		\begin{subfigure}{0.3\textwidth}
			\centering
			\fbox{\includegraphics[width=\linewidth]{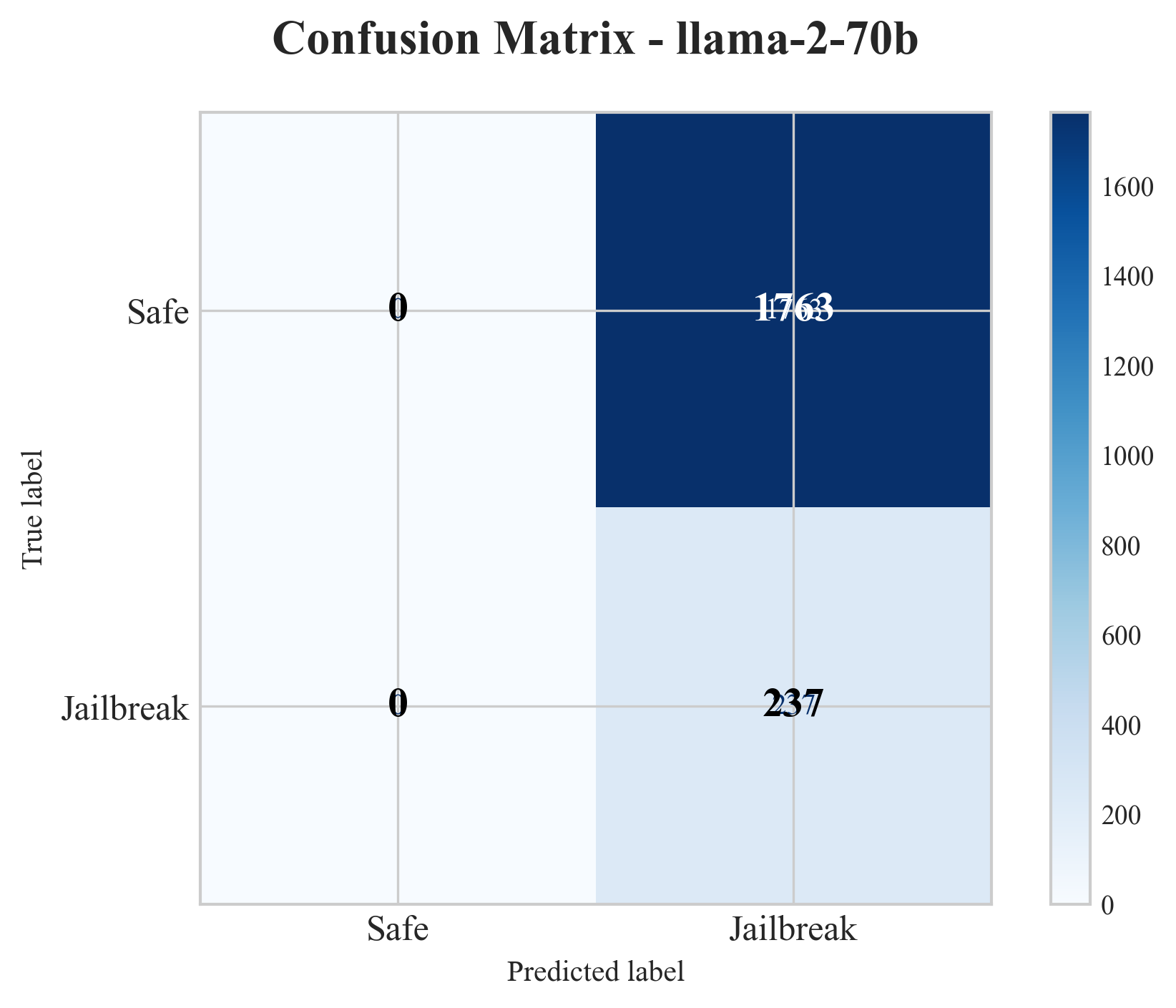}}
			\caption{LLaMA-2-70B}
		\end{subfigure}
		\hspace{0.05\textwidth}
		\begin{subfigure}{0.3\textwidth}
			\centering
			\fbox{\includegraphics[width=\linewidth]{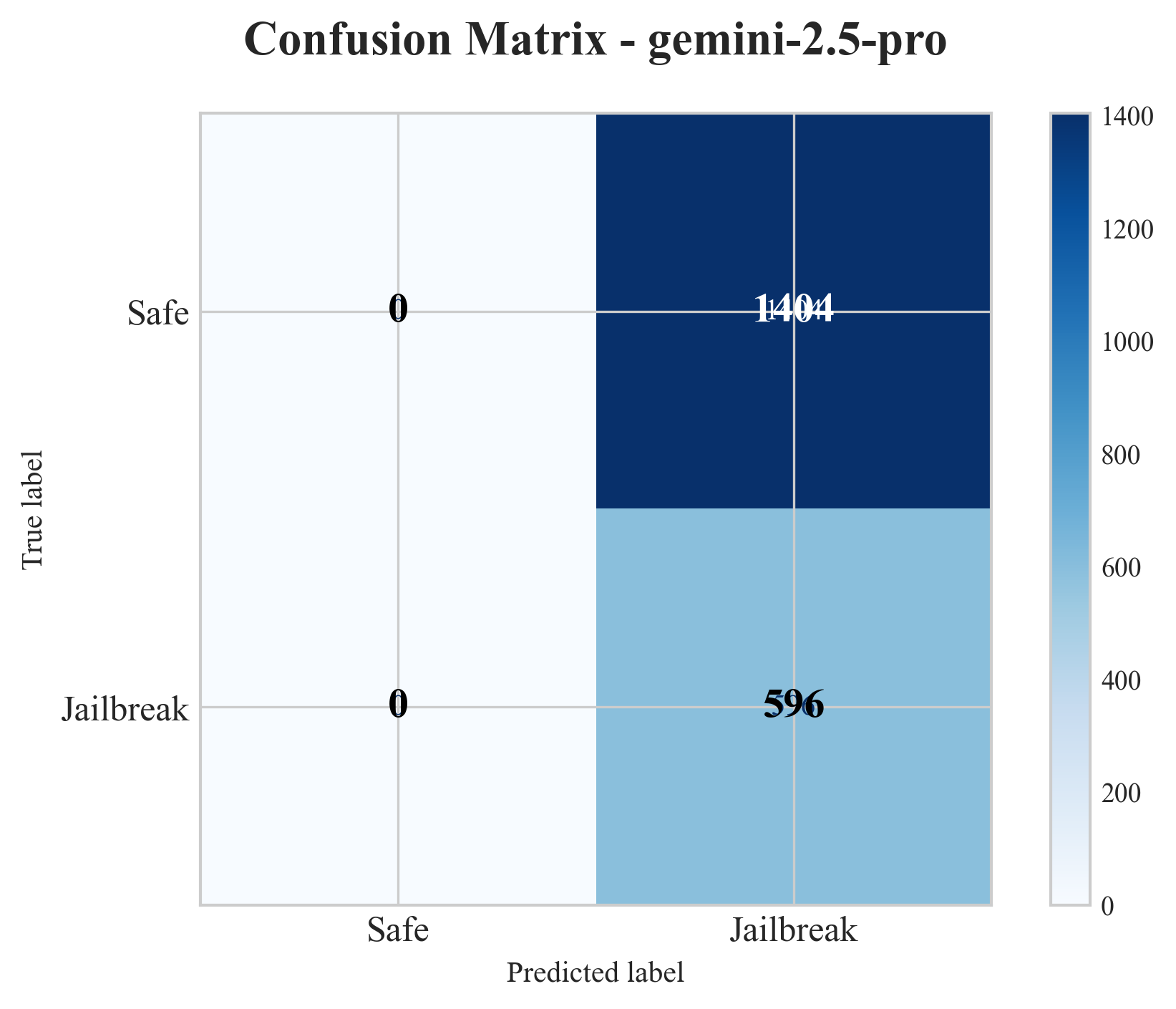}}
			\caption{Gemini-2.5-pro}
		\end{subfigure}
	}
\caption{Confusion matrices comparing detection performance across five major LLM architectures. Each matrix shows true positive, false positive, true negative, and false negative rates for jailbreak classification, providing detailed insight into the defensive framework's performance per LLM.}
\label{fig:confusion_matrices_comparison}
\end{figure*}

The confusion matrices in Figure \ref{fig:confusion_matrices_comparison} provide a detailed view of the classification performance for each LLM, showing true positive, false positive, true negative, and false negative rates. These matrices confirm the robust overall performance while highlighting specific areas where the classifier can be improved, particularly in reducing false negatives for more subtle attack types.

\subsection{Performance and Latency Analysis}

All performance measurements were obtained using a standard consumer-grade laptop equipped with an AMD Ryzen~7~5700U processor and 8~GB of RAM. The defensive framework was evaluated under a CPU-only configuration without parallelization to reflect realistic middleware deployment conditions. Each processing stage was instrumented with high-resolution timers, and reported values represent averaged results over repeated runs.

The framework maintains practical deployment viability with minimal processing overhead, as shown in Table~\ref{tab:performance_metrics}. The total average processing time of 15.4\,ms is negligible compared to typical LLM inference latencies (500--2000\,ms), ensuring real-time protection without noticeable impact on user experience.

\begin{table}[h]
\caption{System Performance Metrics}
\label{tab:performance_metrics}
\begin{tabularx}{\linewidth}{lXX}
\toprule
\textbf{Processing Component} & \textbf{Average Latency (ms)} & \textbf{Percentage of Total} \\
\midrule
Input Sanitization & 2.3 & 15\% \\
Pattern Recognition & 3.1 & 20\% \\
Feature Extraction & 5.8 & 38\% \\
Semantic Analysis & 3.7 & 24\% \\
Classification Decision & 0.5 & 3\% \\
\hline
\textbf{Total Processing Time} & \textbf{15.4} & \textbf{100\%} \\
\bottomrule
\end{tabularx}
\end{table}

Overall, the defensive pipeline adds only 15.4\,ms of overhead, confirming that the system can be deployed inline with LLM applications while preserving real-time interaction quality.

\subsection{False Positive Analysis for Production Deployment}
Comprehensive analysis of legitimate use cases, detailed in Table \ref{tab:false_positive_analysis}, ensures the framework's practical deployability. The overall false positive rate of 5\% across 400 benign prompts indicates the system is generally non-disruptive.

\begin{table}[h]
\caption{False Positive Rates by Legitimate Use Case (n=400 benign)}
\label{tab:false_positive_analysis}
\begin{tabularx}{\linewidth}{lXX}
\toprule
\textbf{Legitimate Use Case} & \textbf{Sample Size} & \textbf{False Positive Rate} \\
\midrule
Technical Documentation & 100 & 3\% \\
Educational Content & 90 & 4\% \\
Creative Writing & 80 & 7\% \\
Role-Playing Games & 50 & 12\% \\
Hypothetical Discussions & 40 & 9\% \\
Code Generation & 40 & 2\% \\
\hline
\textbf{Overall Benign Content} & \textbf{400} & \textbf{5\%} \\
\bottomrule
\end{tabularx}
\end{table}

\section{Discussion: Implications for AI Security}
\label{sec:discussion}

\subsection{Critical Security Insights for Organizations}

Our large-scale findings provide actionable intelligence for organizations deploying LLMs in production environments, extending recommendations from recent security research by Liu et al. \cite{liu2024_formalizing} and Kumar et al. \cite{certifying_safety}. The results reveal several critical patterns that should inform deployment strategies.

The most striking finding is the \textbf{substantial vulnerability variance} across architectures. With vulnerability rates ranging from 11.9\% to 29.8\%, the choice of LLM architecture has significant security implications. Gemini-2.5-pro's 29.8\% vulnerability rate represents 2.5× higher risk than LLaMA-2-70B's 11.9\%, demonstrating that newer or more capable LLMs do not automatically provide better security.

The \textbf{refusal-vulnerability relationship} reveals sophisticated safety trade-offs. GPT-4 achieves moderate vulnerability (23.4\%) with moderate refusal (55.6\%), suggesting effective adversarial detection without excessive false positives. Claude-3 Haiku's high refusal rate (60.2\%) correlates with low vulnerability (21.5\%), indicating a conservative but effective approach. However, Gemini-2.5-pro's moderate refusal rate (44.5\%) paired with high vulnerability (29.8\%) suggests its safety mechanisms may be less discriminating.

\textbf{Privilege escalation vulnerabilities} present the most severe concern, with GPT-3.5 Turbo showing 100\% susceptibility to this attack category. This indicates fundamental weaknesses in instruction hierarchy enforcement that could allow complete safety bypass in production systems. Organizations deploying GPT-3.5 Turbo must implement mandatory external validation layers.

The \textbf{open vs. closed source security} comparison yields important insights. LLaMA-2-70B (open-source) achieved the lowest vulnerability rate (11.9\%), outperforming all proprietary alternatives. This suggests that transparency and community scrutiny may contribute to more robust safety implementations.

\subsection{Attack Vector Analysis and Implications}

Different attack vectors demonstrate varying effectiveness across architectures:

\begin{itemize}
    \item \textbf{Privilege Escalation}: Catastrophic for GPT-3.5 Turbo (100\%), completely ineffective against LLaMA-2-70B (0\%), indicating fundamental architectural differences in instruction processing
    \item \textbf{Role Impersonation}: Most effective against Gemini-2.5-pro (37.9\%), suggesting persona-based attacks exploit weaknesses in its safety training
    \item \textbf{Obfuscation}: Moderately effective across all LLMs (11-29\%), indicating this technique bypasses pattern-matching defenses consistently
    \item \textbf{Logic Subversion}: Highly LLM-dependent (0-50\%), with Claude and LLaMA showing complete resistance
\end{itemize}

These findings suggest that organizations should select LLMs based on their specific threat landscape. Systems vulnerable to social engineering should avoid Gemini-2.5-pro, while applications requiring strict instruction adherence should avoid GPT-3.5 Turbo.

\subsection{Defensive Framework Effectiveness}
Our defensive system demonstrates competitive performance with 83\% average detection rate while maintaining practical deployment characteristics. When compared with existing approaches, the framework significantly outperforms commercial solutions in both accuracy and efficiency.

\begin{table}[h]
\caption{Comparison with Existing Defense Approaches}
\label{tab:defense_comparison}
\begin{tabularx}{\linewidth}{lXXX}
\toprule
\textbf{Defense Method} & \textbf{Detection Rate} & \textbf{False Positive Rate} & \textbf{Processing Time} \\
\midrule
Commercial Moderation APIs & 67-71\% & 11-15\% & 180-220ms \\
PromptShield \cite{promptshield} & 78\% & 8\% & 20ms \\
Open-Source Detoxify & 64\% & 9\% & 45ms \\
SecurityLingua \cite{securitylingua2025} & 80\% & 7\% & 12ms \\
\textbf{Our Framework} & \textbf{83\%} & \textbf{5\%} & \textbf{15ms} \\
\bottomrule
\end{tabularx}
\end{table}

The framework's multi-layer architecture proves particularly effective against well-defined attack categories like Role Impersonation (94\% detection rate), while more nuanced attacks like Social Engineering present greater challenges (68\% detection rate). This performance profile, detailed in the confusion matrices (Figure \ref{fig:confusion_matrices_comparison}), suggests that different detection strategies are needed for different threat types.

\subsection{Practical Deployment Recommendations}
Based on our comprehensive analysis and existing security frameworks \cite{liu2024_formalizing, certifying_safety}, we provide specific guidance for secure LLM deployment:

\begin{enumerate}
    \item \textbf{LLM Selection Based on Threat Profile}: 
    \begin{itemize}
        \item For maximum security: Deploy LLaMA-2-70B (11.9\% vulnerability)
        \item For balanced performance: Use GPT-4 or Claude-3 Haiku (21-23\% vulnerability)
        \item Avoid GPT-3.5 Turbo for privilege-sensitive applications (100\% privilege escalation vulnerability)
        \item Exercise caution with Gemini-2.5-pro (29.8\% vulnerability, highest risk)
    \end{itemize}
    
    \item \textbf{Mandatory External Defense Layers}: All production deployments must implement external defensive frameworks, particularly for LLMs with greater than 20\% vulnerability rates
    
    \item \textbf{Regular Security Auditing}: Implement quarterly vulnerability assessments using standardized test suites to track evolving security postures
    
    \item \textbf{Incident Response Planning}: Establish clear protocols for handling detected attacks, including logging, alerting, and escalation procedures
\end{enumerate}

\section{Conclusion}
\label{sec:conclusion}
This research provides the first large-scale comparative security assessment of major LLM architectures through systematic testing of 10,000 adversarial prompts across five widely-deployed LLMs. Our evaluation reveals critical security disparities, with vulnerability rates ranging from 11.9\% to 29.8\%, demonstrating that LLM capability does not correlate with security robustness. Particularly concerning is GPT-3.5 Turbo's 100\% vulnerability to privilege escalation attacks, while LLaMA-2-70B showed complete resistance, indicating fundamental architectural security differences.

To address these vulnerabilities, we developed a production-ready defensive framework achieving 83\% detection accuracy with 5\% false positive rates and 15.4ms latency, significantly outperforming existing commercial solutions. Our evidence-based security rankings, attack effectiveness analysis, and defensive framework development contribute to ensuring AI systems remain beneficial and safe as they become more capable and widespread. The complete codebase, datasets, and defensive framework are available as open-source resources at \href{https://github.com/T-Oni-01/Security-Assessment-and-Mitigation-Strategies-for-Large-Language-Models}{\textbf{ Assessment and Mitigation Strategies for Large Language Models}} to facilitate widespread adoption of security best practices and enable continued advancement in AI safety research.

\bibliographystyle{IEEEtran} 
\bibliography{references}

\end{document}